\documentclass{IEEEtran}
\usepackage{newlfont}
\usepackage{multicol}
\usepackage{multirow}
\usepackage[T1]{fontenc}
\usepackage[utf8]{inputenc}
\usepackage[fleqn]{amsmath}
\usepackage{amsfonts,amssymb}
\usepackage{xspace}
\usepackage{color}
\usepackage{listings}
\usepackage{newlfont}
\usepackage{url}
\usepackage[noend]{algorithmic}
\usepackage{algorithm}
\usepackage{float}
\usepackage[table]{xcolor}
\usepackage{graphicx}
\usepackage{fancyhdr}
\usepackage{alltt}
\usepackage{caption}
\usepackage{subcaption}
\usepackage{srcltx}
\usepackage{theorem}
\usepackage{latexsym}
\usepackage{textcomp}
\usepackage{setspace}
\usepackage{array}
\usepackage{mdwmath}
\usepackage{mdwtab}
\usepackage{enumerate}
\usepackage{wrapfig}
\usepackage{tikz}
\usepackage{proof}
\usetikzlibrary{positioning,fit,backgrounds,calc,shapes.multipart}
\usepackage{tabularx}
\usepackage{booktabs}
\usepackage{mathtools}
\usepackage{bbm}
\usepackage{stmaryrd}
\usepackage{makecell}
\usepackage{mathpartir}
\usepackage{esvect}
\usepackage{mdframed}
\usepackage{bbold}

\usepackage[colorinlistoftodos,textwidth=4cm, shadow]{todonotes}
\newcommand{\ignore}[1]{}

\definecolor{midgrey}{rgb}{0.3,0.3,0.3}
\definecolor{darkred}{rgb}{0.7,0.1,0.1}
\newcommand{\kuba}[1]{$[$\textcolor{violet}{Jakub}:~~\emph{\textcolor{midgrey}{#1}}$]$}
\newcommand{\nham}[1]{$[$\textcolor{orange}{Nh\^{a}m}:~~\emph{\textcolor{midgrey}{#1}}$]$}
\newcommand{\arie}[1]{$[$\textcolor{darkred}{Arie}:~~\emph{\textcolor{midgrey}{#1}}$]$}
\newcommand{\jorge}[1]{$[$\textcolor{blue}{Jorge}:~~\emph{\textcolor{midgrey}{#1}}$]$}

\newcommand{\ag}[1]{\arie{#1}}

\definecolor{commcolour}{rgb}{0.25,0.25,0.5}

\lstnewenvironment{clisting}{\lstset{language=C}}{}
\lstnewenvironment{cpluslisting}{\lstset{language=C++}}{}

\newcounter{proglineno}

\newcommand{\cpp}{C\texttt{++}\xspace}
\newcommand{\Cpp}{\cpp}
\newcommand{\llvm}{\textsc{LLVM}\xspace}

\newcommand{\publicseahorn}{\textsc{SeaHorn}\xspace}
\newcommand{\seahorn}{\publicseahorn}

\newcommand{\clang}{\textsc{Clang}\xspace}
\newcommand{\SeaDsa}{\textsc{SeaDsa}\xspace}

\newcommand{\Dsa}{\textsc{DSA}\xspace}

\newcommand{\TeaDsa}{\textsc{TeaDsa}\xspace}

\newcommand{\SVF}{\textsc{SVF}\xspace}

\newcommand{\icon}{\textsc{Icon}\xspace}

\newlength{\storetabcolsep}
\AtBeginDocument{
  \let\storearraystretch\arraystretch
  \setlength{\storetabcolsep}{\tabcolsep}
}

\newcommand{\restoretablespace}{
  \let\arraystretch\storearraystretch
  \setlength{\tabcolsep}{\storetabcolsep}}

\newcommand{\code}[1]{\texttt{#1}}
\newcommand{\at}{\mathbin{@}}

\newcommand{\ptsTo}{\mapsto}
\newcommand{\ptsToInM}[1]{\xmapsto{#1}}
\newcommand{\ptsToIn}[1]{\ptsToInM{\code{#1}}}
\newcommand{\ptsToInAtM}[2]{\xmapsto{#1 \at #2}}
\newcommand{\ptsToInAt}[2]{\ptsToInAtM{\code{#1}}{#2}}
\newcommand{\ptsToTy}[1]{\xmapsto{#1}}

\newcommand{\squeezeRule}{\hspace{-2.0em}}
\newcommand{\mkRule}[3][]{\scalebox{0.7}{\inferrule{#2}{#3}{\phantom{a}\textsc{#1}}\hspace{-2.0em}}}

\newcommand{\mkSpace}{\hspace{-0.7em}}

\newcommand{\mkFn}[1]{\mathit{#1}}
\newcommand{\fnOf}[1]{\mathit{fun}(#1)}

\newcommand{\rr}{\code{r}}
\newcommand{\rp}{\code{p}}

\newcommand{\tyT}{\code{T}}

\newcommand{\tyU}{\code{U}}

\newcommand{\tyX}{\code{X}}

\newcommand{\mkVec}[1]{\overline{\mbox{#1\vphantom{z}}}}
\newcommand{\rzk}[0]{\code{z}_k}
\newcommand{\ryk}[0]{\code{y}_k}
\newcommand{\rxk}[0]{\code{x}_k}

\newcommand{\rfk}[0]{\code{f}_k}

\newcommand{\fldA}{\code{a}}
\newcommand{\fldB}{\code{b}}

\newcommand{\cvr}{$\mkVec{\rr}$\xspace}
\newcommand{\cvx}{$\mkVec{\code{x}}$\xspace}
\newcommand{\cvy}{$\mkVec{\code{y}}$\xspace}
\newcommand{\cvz}{$\mkVec{\code{z}}$\xspace}

\newcommand{\cvf}{$\mkVec{\code{f}}$\xspace}
\newcommand{\cvp}{$\mkVec{\rp}$\xspace}

\newcommand{\fnDecl}[1]{\code{fun #1(\cvf): \cvr}}

\newcommand{\sibling}{\mkFn{siblingObj}}
\newcommand{\isFormal}[1]{\mkFn{isFormal}(#1)}

\newcommand{\resolve}{\mkFn{Resolve}}

\newcommand{\accessible}{\mkFn{Accessible}}
\newcommand{\fldOf}[1]{\mkFn{fld}(#1)}

\newcommand{\caller}{\code{caller}}
\newcommand{\callee}{\code{callee}}

\newcommand{\tyChar}[1]{\code{char#1}}
\newcommand{\tyInt}[1]{\code{int#1}}
\newcommand{\tyFloat}[1]{\code{float#1}}

\newtheorem{theorem}{Theorem}

\renewcommand{\paragraph}[1]{\noindent\textbf{#1.}}
\title{Unification-based Pointer Analysis without Oversharing}

\author{
  \IEEEauthorblockN{Jakub Kuderski\IEEEauthorrefmark{1},
                    Jorge A. Navas\IEEEauthorrefmark{2}, and
                    Arie Gurfinkel\IEEEauthorrefmark{1}}\\
  \IEEEauthorblockA{\IEEEauthorrefmark{1}University of Waterloo, Canada
    \\\{jakub.kuderski, arie.gurfinkel\}@uwaterloo.ca}\\
  \IEEEauthorblockA{\IEEEauthorrefmark{2}SRI International, USA
    \\jorge.navas@sri.com}
}

\begin{document}
\maketitle

\begin{abstract}
  Pointer analysis is indispensable for effectively verifying
  heap-manipulating programs. Even though it has been studied
  extensively, there are no publicly available pointer analyses that
  are moderately precise while scalable to large real-world
  programs. In this paper, we show that existing context-sensitive
  unification-based pointer analyses suffer from the problem of
  \emph{oversharing} -- propagating too many abstract objects across
  the analysis of different procedures, which prevents them from scaling to
  large programs. We present a new pointer analysis for \llvm, called \TeaDsa,
  without such an oversharing. We show how to further improve precision and
  speed of \TeaDsa with extra contextual information, such as flow-sensitivity
  at call- and return-sites, and type information about memory accesses. We
  evaluate \TeaDsa on the verification problem of detecting unsafe memory
  accesses and compare it against two state-of-the-art pointer analyses: \SVF
  and \SeaDsa. We show that \TeaDsa is one order of magnitude faster than either
  \SVF or \SeaDsa, strictly more precise than \SeaDsa, and, surprisingly,
  \mbox{sometimes more precise than \SVF.}
  
\end{abstract}

\section{Introduction}
\label{sec:intro}

Pointer analysis (PTA) -- determining whether a given pointer aliases with
another pointer (\emph{alias analysis}) or points to an allocation site
(\emph{points-to analysis}) are indispensable for reasoning about low-level code
in languages such as C, \Cpp, and \llvm bitcode. In compiler optimization, PTA is
used to detect when memory operations can be lowered to scalar operations and
when code transformations such as code motion are sound. In verification and
bug-finding, PTA is often used as a pre-analysis to limit the implicit
dependencies between values stored in memory. This is typically followed by
a deeper, more expensive, path-sensitive
analysis~(e.g., \cite{DBLP:conf/icse/YanSCX18,pinpoint,seahorn}).
In both applications, the efficiency of PTA is crucial since
it directly impacts compilation and verification times, while precision of the
analysis determines its usability. Moderately precise and efficient PTA is
most useful, compared to precise but inefficient or efficient but imprecise
variants.

The problem of pointer analysis is well studied. A survey by
Hind~\cite{introduction-hind} (from~2001!) provides a good overview of
techniques and precision vs cost trade-offs. Despite that, very few practical
implementations of PTA targeting low-level languages 
are available. In part, this is explained by the difficulty of soundly
supporting languages that do not provide memory safety guarantees, allow
pointers to fields of aggregates, and allow arbitrary pointer arithmetic.
In this paper, we focus on the PTA problem for low-level languages.

There are many dimensions that affect precision vs cost trade-offs of
a PTA, including path-, flow-, and (calling) context-sensitivity,
modeling of aggregates, and modularity of the analysis. From the
efficiency perspective, the most significant dimension is whether the
analysis is \emph{inclusion-based} (a.k.a.,
\emph{Andersen-style}~\cite{Andersen1994ProgramAA}) or
\emph{unification-based} (a.k.a.,
\emph{Steensgaard-style}~\cite{Steensgaard96AA}). All other things
being equal, a unification-based analysis is significantly faster than
an inclusion-based one at the expense of producing very imprecise
results. To improve further precision while retaining its efficiency,
a unification-based PTA can be extended with (calling)
context-sensitivity in order to separate local aliasing created at
different call sites. Unfortunately, the combination of a
unification-based analysis with context-sensitivity can quickly
degenerate in a prohibitive analysis.

State-of-the-art implementations of unification-based, context-sensitive PTA
(e.g., \Dsa~\cite{dsa} and \SeaDsa~\cite{sea-dsa}) perform the analysis in
phases. First, each function is analyzed in an intra-procedural manner
(\textsc{Local}). Second, a \textsc{Bottom-Up} phase inlines callees' points-to
graphs into their callers. Third, a \textsc{Top-Down} phase inlines callers'
points-to graphs into their callees. We observed that both \textsc{Bottom-Up}
and \textsc{Top-Down} often copy too many \emph{foreign objects}, memory objects
allocated by other functions that cannot be accessed by the function at hand,
increasing dramatically both analysis time and memory usage. In fact, we show in
Sec.~\ref{sec:evaluation} that the majority of analysis runtime is spent on
copying foreign objects. Even worse, due to the imprecise nature of
unification-based PTA and difficulty of analyzing accurately aggregates, foreign
objects can be aliased with other function objects affecting negatively the
precision of the analysis. We refer to \emph{oversharing} as the existence of
large number of inaccessible \mbox{foreign objects during the analysis of a particular
function.}

In this paper, 
we present a new pointer analysis for \llvm, called \TeaDsa,
that eliminates a class of such an oversharing.
\TeaDsa is a new unification-based PTA implemented on top of \SeaDsa.
Since \TeaDsa builds on \SeaDsa, it remains modular (i.e., analysis of each
function is summarized and the summary is used at call sites), context-, field-,
and array-sensitive. The first main difference is that \TeaDsa does not
add oversharing during \textsc{Top-Down} while retaining full
context-sensitivity. This is achieved by not copying foreign objects
coming from callers.
This is a major improvement compared to previous implementations. \Dsa
mitigates the oversharing problem by partially losing
context-sensitivity. \SeaDsa does not tackle this problem 
 since it focuses on medium-size programs such as
the SV-COMP benchmarks~\cite{svcomp19}.

Second, we observed that oversharing can also come from the \textsc{Local} phase.
This is mainly because the local analysis is flow-insensitive. To mitigate
this, we make \TeaDsa flow-sensitive but only at call- and return
sites. This preserves the efficiency of the analysis while improving its
precision.

Third, we noted that another source of imprecision in \SeaDsa is
loss of field-sensitivity during analysis of operations in which
determining the exact field being accessed is difficult and merging
that is inherent to its unification nature. Crucially, in many cases
where field-sensitivity is lost, it is still clear that pointers do
not alias if their types are taken into account. Under \emph{strict
  aliasing} rules of the C11 standard, two pointers cannot alias if
they do not have compatible types \cite{ISO:2011:IIIb}. By following strict
aliasing, we further improve the precision of \TeaDsa.

We have evaluated \TeaDsa against \SeaDsa and \SVF, a state-of-the-art
inclusion-based pointer analysis in \llvm, on the verification problem
of detecting unsafe memory accesses. Our evaluation shows that \TeaDsa
is one order of magnitude faster than \SeaDsa or \SVF, strictly more
precise than \SeaDsa, and sometimes more precise than \SVF.

 \section{Overview}
\label{sec:overview}
\newcommand{\refsubfig}[2]{Fig.~\ref{#1}\subref{#2}}

\begin{figure*}[t]
  \centering
  \hspace*{2em}
  \begin{subfigure}[b]{0.60\linewidth}
    \begin{lstlisting}[language=C,numbers=left,multicols=2,
                       basicstyle=\scriptsize\ttfamily,
                       keywordstyle=\color{blue}\ttfamily,
                       stringstyle=\color{brown}\ttfamily]
const char *str1 = "Str1";  <@\abstractObject{1}@>
const char *str2 = "Str2";  <@\abstractObject{2}@>
const char *str3 = "Str3";  <@\abstractObject{3}@>

void print(const char *x) {}

const char *getStr() {
  const char *p = nondet() ?
                  str1 : str2;
  print(p);
  return str1;
}

struct Config
{ const char *label; int *val; };
int foo(struct Config *conf) {
  const char str4[5] = "Str4";<@\abstractObject{4}@>
  print(str4);
  const char *r = getStr();
  print(r);
  return conf->label == r;
}

int bar() {
  int i = 42;                 <@\abstractObject{5}@>
  const char *s = nondet() ?
                  str2 : str3;
  struct Config c = {s, &i}   <@\abstractStruct{u}@>
  return foo(&c);
}
    \end{lstlisting}
    \caption{\label{fig:p1_listing}}
  \end{subfigure}
  \hspace*{-1.7em}
  \begin{subfigure}[b]{0.36\linewidth}
    \centering\includegraphics[width=170pt]{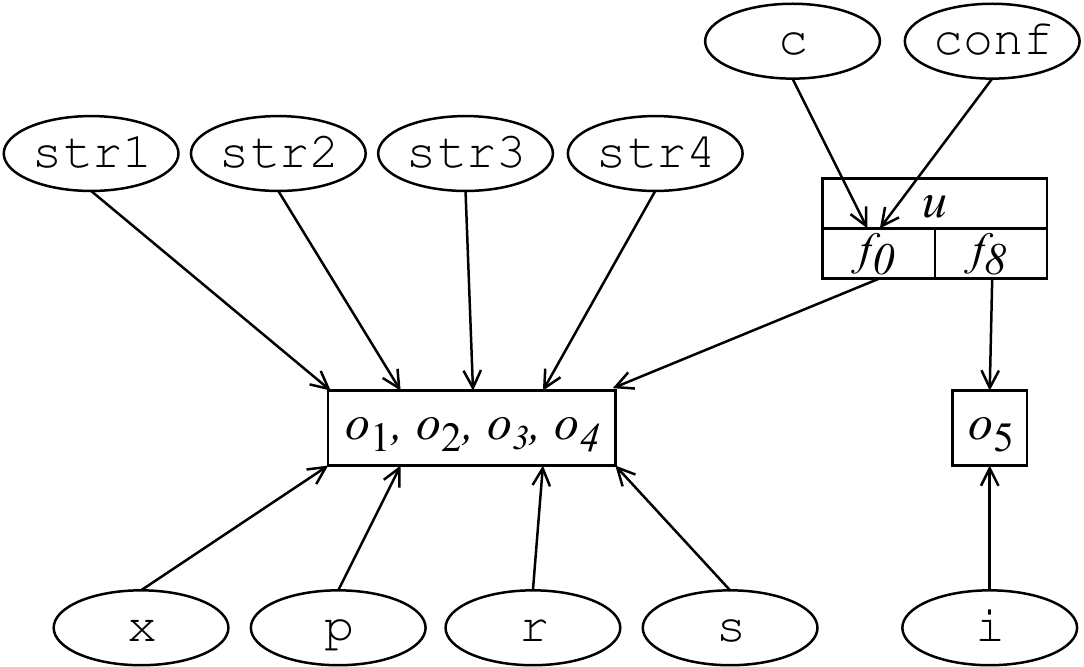}
    \caption{\label{fig:p1_steens}}
  \end{subfigure}
  \vspace{0.6em}
  \caption{Sample C program $P_1$ \subref{fig:p1_listing} and its
           Context-insensitive Points-To Graph $G_1$~\subref{fig:p1_steens}.\label{fig:ov_p1}}
\end{figure*}

In this section, we illustrate our approach on a series of simple examples.
Consider a C program $P_1$ in \refsubfig{fig:ov_p1}{fig:p1_listing} and its
corresponding context-insensitive and flow-insensitive \mbox{points-to} graph $G_1$ in
\refsubfig{fig:ov_p1}{fig:p1_steens}. The nodes of $G_1$ correspond to
registers (ellipses) and \emph{groups} of abstract memory objects (rectangles),
and edges of $G_1$ represent the points-to relation between them.
As usual, a \emph{register} is a program variable whose 
address is not taken. For example, the local variable \texttt{s} is a register.
Similarily, an \emph{abstract object} represents concrete memory objects allocated
at a static allocation site, such as an address-taken global or local variable,
or a call to an allocating function like \texttt{malloc}.
For field sensitivity, \texttt{struct} fields are associated with their own abstract objects.
In \refsubfig{fig:ov_p1}{fig:p1_listing}, we denote corresponding abstract objects in comments.
For example, the local integer variable \texttt{i} is associated with an abstract object $o_5$,
while the \texttt{struct} variable \texttt{c} is associated with abstract
objects $u.f_0$ and $u.f_8$ for its \texttt{label} 
and \texttt{val} fields at offset~$0$ and~$8$, respectively.

The edges of $G_1$ denote whether a pointer $p$ may point to an abstract object $o$, written 
$p \mapsto o$. Whenever $p$ may point to multiple abstract objects all of these
objects are grouped into a single (rectangular) node.
For instance, $\texttt{x} \mapsto o_1$, $\texttt{x} \mapsto o_2$, $\texttt{x}
\mapsto o_3$, $\texttt{x} \mapsto o_4$, or $\texttt{x} \mapsto \{o_1, o_2, o_3,
o_4\}$ for brevity.
We say that two pointers $p_1$ and $p_2$ \emph{alias} when they may point
to the same abstract object, written $\mathit{alias}(p_1, p_2)$. 

\lstset{escapeinside={<@}{@>}}
\newcommand{\abstractObject}[1]{\color{magenta}// $\mathbf{o_{#1}}$}
\newcommand{\abstractStruct}[1]{\color{magenta}// $\mathbf{#1}$}

The graph $G_1$ in \refsubfig{fig:ov_p1}{fig:p1_steens} corresponds to the Steensgaard
(or unification-based) PTA~\cite{Steensgaard96AA}. This style of PTA ensures an
invariant (\textbf{I1}): whenever there is a pointer $p_1$ and objects $o_a$
and $o_b$ such that $p_1 \mapsto o_a$ and $p_1 \mapsto o_b$, then for any other
pointer $p_2$ if $p_2 \mapsto o_a$ then $p_2 \mapsto o_b$. On one hand,
(\textbf{I1}) implies that Steensgaard PTA can be done in linear time using a
union-find data structure to group objects together. On the other, Steensgaard
PTA is quite imprecise. In our running example, it deduces that almost all
registers of $P_1$ may alias, which is clearly not the case. For instance,
$\texttt{s} \mapsto \{o_1, o_2, o_3, o_4\}$ in \refsubfig{fig:ov_p1}{fig:p1_steens},
even though there is no execution in which $\code{s} \ptsTo o_1$ or $\code{s}
\ptsTo o_4$.

\begin{figure*}[t]
  \centering
  \hspace*{1.0em}
  \begin{subfigure}[b]{0.48\linewidth}
    \centering\includegraphics[width=255pt]{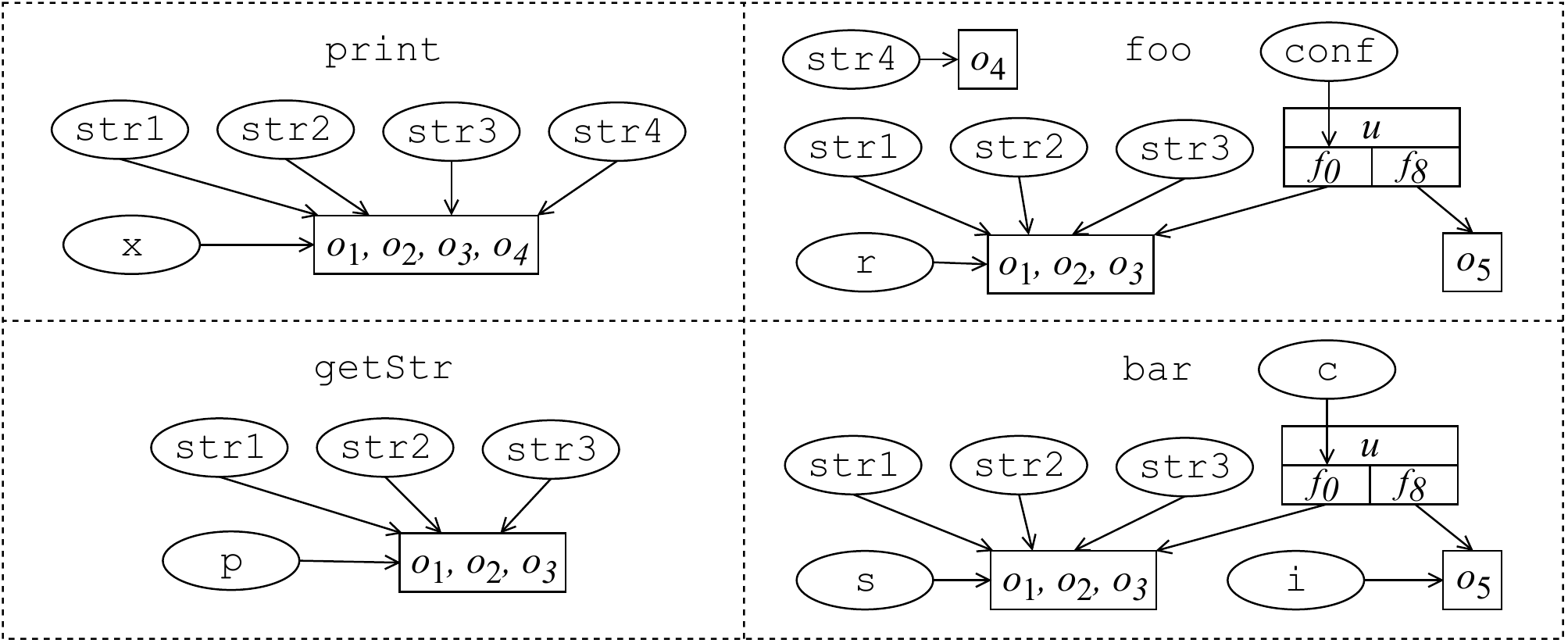}
    \caption{\label{fig:p1_dsa}}
  \end{subfigure}
  \begin{subfigure}[b]{0.48\linewidth}
    \centering\includegraphics[width=235pt]{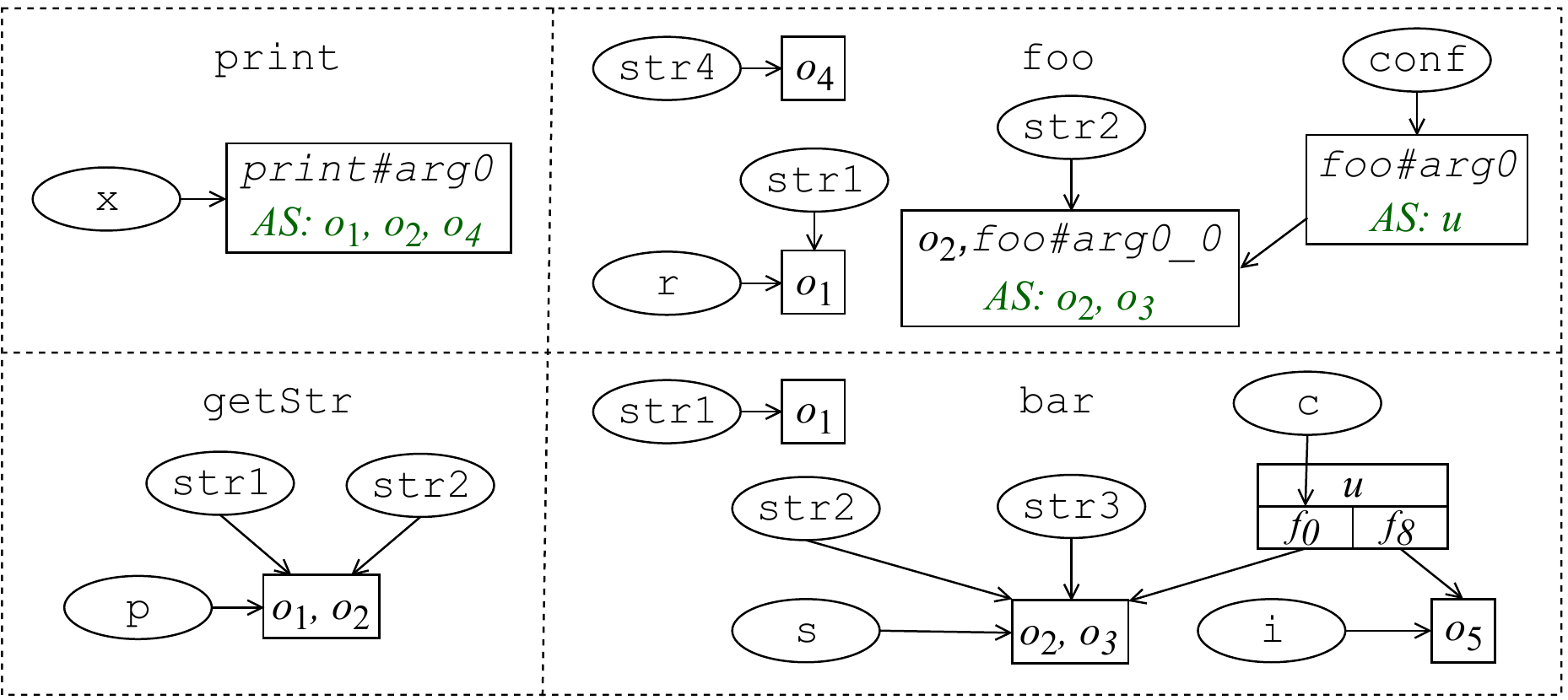}
    \caption{\label{fig:p1_teadsa}}
  \end{subfigure}
  \caption{Context-sensitive Points-To Graphs for $P_1$.
           \label{fig:p1_cs_steens}}
\end{figure*}

A standard way to make the Steensgaard PTA more precise is to perform the
analysis separately for each procedure. This is referred to as
(calling) context-sensitivity. The main idea is to distinguish local aliasing
created at different call sites. Data Structure Analysis (\Dsa) \cite{dsa} is an
example of a context-sensitive Steensgaard PTA. The results of a
context-sensitive Steensgaard PTA on $P_1$ are shown in
\refsubfig{fig:p1_cs_steens}{fig:p1_dsa} as four separate points-to graphs --
one for each procedure in $P_1$. An increase in precision (compared to the PTA
in \refsubfig{fig:ov_p1}{fig:p1_steens}) is visible in procedures \texttt{foo},
\texttt{bar}, and \texttt{getStr}: the string \texttt{str4} does not alias all
the other strings. The improvement comes at a cost -- some abstract
objects appear in the analysis results of multiple procedures. For instance,
$o_1$, $o_2$, and $o_3$ appear in all 4 graphs. In the worst case, \Dsa
can grow quadratically in the program size, which prevents it from
scaling to large programs.

In Sec.~\ref{sec:evaluation}, we show that in \Dsa the majority of runtime is often
spent on copying \emph{foreign abstract objects} coming from other procedures. For
example, consider the abstract objects $u.f_8$ and $o_5$: the procedure
\texttt{foo} never accesses the \texttt{val} field of \texttt{conf}. As shown in
\refsubfig{fig:ov_p1}{fig:p1_listing}, $u.f_8$ and $o_5$ are only accessible in
\texttt{foo} through \texttt{conf} and thus should not appear in the analysis
for \texttt{foo} or any of its callees. However, both $u.f_8$ and $o_5$ are
present in the points-to graph for \texttt{foo} in
\refsubfig{fig:p1_cs_steens}{fig:p1_dsa}, as computed by a DSA-like PTA. This
performance issue was already observed in~\cite{dsa}, but only a workaround that
loses context-sensitivity for global objects was implemented.

In this paper, we show that points-to analysis should refer only to abstract
objects actually used by a procedure. This includes abstract objects in a
procedure and its callees, abstract objects derived from function arguments, and
used global variables. Thus, foreign abstract objects coming from callers are
not only unnecessary in the final analysis results of their callees, but
needless in the first place. Compared to
\refsubfig{fig:p1_cs_steens}{fig:p1_dsa}, in our proposed analysis,
\refsubfig{fig:p1_cs_steens}{fig:p1_teadsa}, function argument accesses are
given separate abstract objects, instead of referring to (foreign) abstract
objects of callees.

Furthermore, we observed that \Dsa maintains the following invariants
(\textbf{I2}):~if a procedure $F_1$ with $p_1 \mapsto o$ calls a procedure $F_2$,
and there is an interprocedural assignment to a function argument $p_2$ of $F_2$,
${p_2 := p_1}$, then $p_2 \mapsto o$;
(\textbf{I3}):~if $F_1$ calls $F_2$ and
$p_2 \mapsto o$ in $F_2$, and there is an interprocedural assignment
to a pointer $p_1$ in $F_1$ by returning $p_2$ from $F_2$, $p_1 := p_2$,
then $p_1 \mapsto o$.
For example, \texttt{foo} calls \texttt{getStr} in \refsubfig{fig:ov_p1}{fig:p1_listing},
$\texttt{str1} \mapsto \{o_1, o_2\}$ in \texttt{getStr}, thus the returned value
${\texttt{r} \mapsto \{o_1, o_2\}}$. (\textbf{I2}) and (\textbf{I3})~
are useful to argue that adding context-sensitivity to Steensgaard
preserves soundness. However, they cause unnecessary propagations of foreign abstract objects.
For instance, even though according to (\textbf{I1}) it must be that
locally ${\texttt{str1} \mapsto \{o_1, o_2\}}$ in \texttt{getStr},
\texttt{getStr} can only return a pointer to $o_1$, as \texttt{str1} is used
in the return statement,
so ${\texttt{r} \mapsto o_1}$ and ${\texttt{r} \not \mapsto o_2}$ 
-- \mbox{that violates (\textbf{I3})}.

In addition to not introducing foreign abstract object for arguments,
many propagations caused by a local imprecision are avoided by not
maintaining (\textbf{I2}) and (\textbf{I3}).
Breaking (\textbf{I2}) allows the analysis to propagate fewer foreign
abstract objects from callers to callees (i.e., top-down), while breaking
(\textbf{I3}) at return sites to reduces the 
number of maintained foreign abstract objects coming from callees (i.e., bottom-up).

In this paper, we show that a context-sensitive unification-based PTA that does
not maintain (\textbf{I2}) and (\textbf{I3}) can be refined with extra
contextual information to reduce the number of foreign abstract
objects, as long as the information is valid for a given source
location in the current calling context.

\begin{figure}
\hbox{\hspace{2em}\begin{lstlisting}[language=C,numbers=left,
                     basicstyle=\scriptsize\ttfamily,
                     keywordstyle=\color{blue}\ttfamily,
                     stringstyle=\color{brown}\ttfamily]
const int INT_TAG = 0, FLOAT_TAG = 1;
typedef struct { int tag; } Element;
typedef struct
{ Element e; int *d; } IElement;
typedef struct
{ Element e; float *d; } FElement;

void print_int(int);
void baz() {
  int a = 1;                       <@\abstractObject{6}@>
  float f;                         <@\abstractObject{7}@>             
  IElement e1 = {{INT_TAG}, &a};   <@\abstractStruct{v}@>
  FElement e2 = {{FLOAT_TAG}, &f}; <@\abstractStruct{w}@>
  Element *elems[2] = {&e1, &e2};  <@\abstractStruct{x}@>

  for (int i = 0; i < 2; ++i)
    if (elems[i]->tag == INT_TAG) {
      IElement *ie = elems[i];
      int *ip = (int *) ie->d;
      print_int(*ip);
    }
}
  \end{lstlisting}}
  \caption{\label{fig:p2_listing} Sample C program $P_2$.}
\end{figure}

\begin{figure}
    \centering\includegraphics[width=140pt]{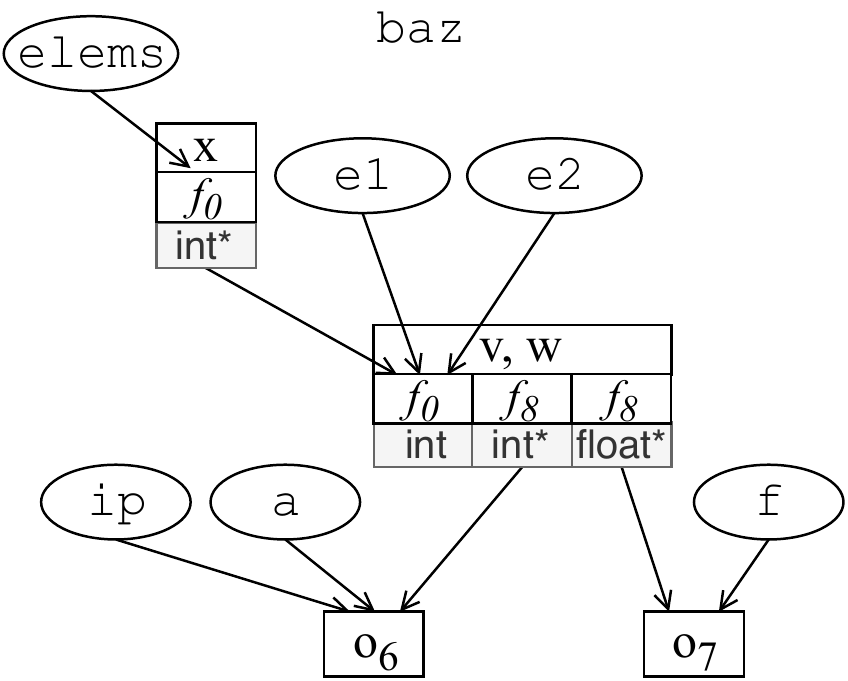}
\caption{Type-aware Points-To Graph of $P_2$.\label{fig:p2_teadsa}}
\end{figure}

The \emph{strict aliasing} rules of the C11  standard
specify that at any execution point every memory location has a type, called
\emph{effective type}. A read from a memory location can only access a type
compatible with its effective type. Consider the program $P_2$ in
Fig.~\ref{fig:p2_listing}: dereferencing the integer pointer
\texttt{ip} is only allowed when the last type written was $\texttt{int}$. We
use \emph{strict aliasing} to improve precision of our PTA.

In order to use types as an additional context, we add an extra
abstract object for any type used with the corresponding allocation
site or its field. As a result, every abstract object has an
associated type tag. Following \emph{strict aliasing}, two objects
$o_1$ and $o_2$ can alias, only when their type tags are compatible.
In the $P_2$'s points-to graph in Fig.~\ref{fig:p2_teadsa}, type tags
are shown at the bottom of each abstract object. We maintain soundness
by discovering type tags based only on memory accesses performed,
instead of relying on casts or type declarations. Alternatively, it is
also possible to use externally supplied type tags \mbox{(e.g.,
  emitted from a C compiler's frontend).}

Although types increase the number of abstract objects, they improve the
precision of our analysis. For example, consider the structs \texttt{e1} and
\texttt{e2} defined in lines 12 and 13 of $P_2$. The \texttt{d} field of
\texttt{e1} is assigned a pointer to \texttt{a}, while the \texttt{d} field of
\texttt{e2} is assigned a pointer to \texttt{f}.
Because of these memory writes, we know that \texttt{e1.d} is of type
\texttt{int*} and \texttt{e2.d} is \texttt{float*}.
Even though $v$ and $w$ are grouped according to (\textbf{I1}) as
${\texttt{elems} \mapsto \{v.f_0.\mathit{int},  w.f_0.\mathit{int}\}}$, $o_6$
and $o_7$ do not alias, as the abstract objects for \texttt{e1.d} and
\texttt{e2.d} differ in type tags: ${v.f_8.\mathit{int*}}$ vs
${w.f_8.\mathit{float*}}$.

In summary, our enhancements to the standard context-sensitive
unification-based PTA not only dramatically improve the performance,
but also the precision of the analysis. This is due to the interaction
between improved local reasoning at call- and return-sites, and the
reduction on propagating foreign abstract objects across functions. We
also show that while the added type-awareness increases the number of
abstract objects, the analysis scales better than a type-unaware one
(on our benchmarks). Interestingly, our proposed PTA is much faster and
usually as precise as the \SVF PTA \cite{svf}, and sometimes even
significantly more precise. Note that SVF is a state-of-the art,
inclusion-based PTA that chooses not to maintain (\textbf{I1}) for
more precision, but is not context-sensitive in order to scale.

 \section{Background}
\label{sec:background}

\newcommand{\ruleSet}[1]{\mathbb{\Gamma}_{\textsc{#1}}}

In this section, we present the necessary background to understand the
rest of the paper. 
We assume a basic understanding of pointer analysis. We
refer interested readers
to~\cite{pointer-analysis-tutorial,introduction-hind} for additional
exposition.

\begin{figure}[t]
  \centering
  \fbox{
    \scriptsize
    \begin{tabular}{lcl}
      $P$   & ::= & ${F+}$ \\
      $F$   & ::= & $\code{fun name(\cvf): \cvr \{} {I+} \code{\}}$ \\
      $I$   & ::= & $\code{r = alloc()} \mid \code{r = cast }T \code{, p} \mid{}$ \\
            &     & $\code{r = load } PT \code{ p} \mid \code{store r, } PT \code{ p} \mid{}$ \\
            &     & $\code{r = gep } PT \code{ p, } \mathit{fld} \mid \code{\cvr = callee(\cvp)} \mid \code{return \cvz}$ \\
      $T$   & ::= & $BT \cup PT$ \\
      $BT$  & ::= & $\tyInt{} \mid \tyFloat{} \mid \tyChar{}$ \\
      $PT$  & ::= & $BT\code{*} \mid BT\code{**}$ \\    
      $\mathit{fld}$ & ::= & $\fldA \mid \fldB$ \\ 
    \end{tabular}
  }
\caption{\label{fig:lang} A simple language.}
\end{figure}

For presentation, we use a simple \llvm-like language shown in 
Fig.~\ref{fig:lang}. The language is used to simplify the presentation, but our
implementation (in Sec.~\ref{sec:evaluation}) supports full \llvm bitcode.
Our language supports standard pointer and memory operations, but has no control
flow constructs, such as conditional statements or loops, and defines a function
by an unordered bag of instructions. This simplified setting is sufficient
because our PTA is flow-insensitive -- it does not use control-flow information.
Although there are no global variables, they are modeled by explicitly passing
them between functions. We allow passing and returning multiple
values, modeled as a vector of function arguments and returns, respectively.
For simplicity of presentation, we assume that all allocations create structures
with exactly two fields, and that the size of an allocation is big enough to
store any scalar type, including
integers and pointers.
New memory objects are created using the \code{alloc} instruction that allocates two
fresh memory objects (one for each field) and returns a pointer to the first
one. The result is saved in a register of type \tyChar{*}, that can be cast to
a desired type with the \code{cast} instruction. Contents of a register is
written to memory using \code{store} and read back with \code{load}. A
\emph{sibling} memory object $I$ of an object $H$ corresponding to field $\fldA$
is obtained with the \code{gep} (\code{GetElementPointer}) instruction with
$\fldB$ as its field operand; applying the \code{gep} to $H$ (or $I$) with a
field operand $\fldA$ yields $H$ (or $I$).

In PTA, the potentially infinite set of concrete memory object is mapped to a
finite set of abstract objects. A standard way to identify abstract objects
is by their \emph{allocation site} -- an \code{alloc} instruction that created
them. 
A \emph{points-to analysis (PTA)} of a program $P$ computes a relation ${\cdot
  \ptsTo \cdot}$, called points-to, between pointers and abstract objects. A PTA 
is sound if whenever $p \not \ptsTo o$ then there is no execution of $P$ in
which $p$ points to a concrete memory object corresponding to $o$. We represent
PTAs using 
inference rules that derive facts of the $\ptsTo$
relation. A PTA is computed by applying these rules until saturation.
Fig.~\ref{fig:rules_inclusion} contains a set of standard inference rules for
the inclusion-based (Andersen-style) context-insensitive analysis in
our language. We let $\ruleSet{I}$ represent the rules in
Fig.~\ref{fig:rules_inclusion} and denote a $\ptsTo$ fact derivable by applying
them exhaustively on a program $P$, written: $\ruleSet{I} \; \vdash_P \; x
\ptsTo H$, where $x$ is a pointer and $H$ is an abstract object. To support the
\code{gep} instruction and make the PTA \emph{field-sensitive}, we extend
$\ruleSet{I}$ with additional rules $\ruleSet{Fld}$ shown in
Fig.~\ref{fig:rules_fields}.

A \emph{unification-based (Steensgaard-style)} PTA is obtained by extending the
analysis with additional unification rules $\ruleSet{U}$ shown
in Fig.~\ref{fig:rules_steens}, such that $\ruleSet{Steens} = \ruleSet{I} \cup
\ruleSet{Fld} \cup \ruleSet{U}$. The rules $\ruleSet{U}$ enforce the invariant
(\textbf{I1}) from Sec.~\ref{sec:overview}. Note that $\ruleSet{Steens}$ is less
precise than $\ruleSet{I} \cup \ruleSet{Fld}$, because altering a PTA by adding
extra inference rules never derives fewer $\ptsTo$ facts. A
\emph{unification-based} PTA like $\ruleSet{Steens}$ is typically implemented using the
\emph{Union-Find} data structure that allows to perform the abstract objects grouping
in (almost) linear time. 

\mprset{vskip=0.5ex}
\begin{figure}
\begin{mdframed}
\begin{mathpar}
  \mkRule[Alloc]{i: \code{r = alloc()}}{\rr \ptsTo H_i}
  
  \mkRule[Cast]{\code{r = cast PT, p} \\ \mkSpace \rp \ptsTo H}{\rr \ptsTo H}
  \\
  \mkRule[Load]{\code{r = load PT p} \\\\ \rp \ptsTo H \\ \mkSpace H \ptsTo I}{\rr \ptsTo I}

  \mkRule[Store]{\code{store r, PT p} \\\\ \rp \ptsTo I \\ \mkSpace \rr \ptsTo H }{I \ptsTo H} 
\end{mathpar}
\end{mdframed}
\caption{Inference rules for Inclusion-based PTA: $\ruleSet{I}$. \label{fig:rules_inclusion}}
\end{figure}

\begin{figure}
\begin{mdframed}
\begin{mathpar}
  \squeezeRule
  \mkRule[GEP]{\code{r = gep PT p, a} \\ \mkSpace \rp \ptsTo H}{\rr \ptsTo H}
  
  \mkRule[GEP]{\code{r = gep PT p, b} \\ \mkSpace \rp \ptsTo H \\\\
               \fldOf{H} = \fldA \\ \mkSpace \sibling(H) = I} 
              {\rr \ptsTo I}
\end{mathpar}
\end{mdframed}
\caption{Inference rules for Field-Sensitivity: $\ruleSet{Fld}$. \label{fig:rules_fields}}           
\end{figure}

\begin{figure}
\begin{mdframed}
\begin{mathpar}
  \mkRule[Incoming]{\rr \ptsTo H \\ \mkSpace \rr \ptsTo I \\\\
                    \phantom{\rp \ptsTo X } \\ \mkSpace \rp \ptsTo I} 
                   {\rp \ptsTo H}
 
  \mkRule[Incoming]{H \ptsTo I \\ \mkSpace H \ptsTo J \\\\
                   \phantom{H \ptsTo I} \\ \mkSpace  L \ptsTo J} 
                   {L \ptsTo I}
 
  \mkRule[Outgoing]{\rr \ptsTo H \\ \mkSpace H \ptsTo J
                    \\\\ \rr \ptsTo I \\ \mkSpace I \ptsTo K} 
                  {H \ptsTo K}
 
 \mkRule[Outgoing]{H \ptsTo I \\ \mkSpace I \ptsTo K
                   \\\\ H \ptsTo J \\ \mkSpace J \ptsTo L} 
                  {I \ptsTo L}

\end{mathpar}
\end{mdframed}
\caption{Unification rules: $\ruleSet{U}$. \label{fig:rules_steens}}
\end{figure}

 \section{Keep Your Objects to Yourself}
\label{sec:rules}

This section is organized as follows: first, we describe how to extend the
$\ruleSet{Steens}$ PTA to be \emph{interprocedural} and explain
\emph{(calling) context-sensitivity}. Next, we show how to extend
$\ruleSet{Steens}$ to a \Dsa-style analysis. Using this formulation, we define the
\emph{oversharing} that happens in \Dsa, and show a way to reduce it. Finally,
we show how to make the PTA partially \emph{flow-sensitive} to further improve
both precision and efficiency.

\paragraph{Context-sensitivity}
The unification-based PTA $\ruleSet{Steens}$ from Sec.~\ref{sec:background} is
an \emph{intraprocedural analysis}. It analyzes a single function at a time and
does not reason about other functions. \emph{Interprocedural} 
reasoning requires propagating $\ptsTo$ between \emph{callers} and
\emph{callees} at all call-sites. For simplicity of explanation, we assume that
calls are direct, i.e., callees are statically known, and that functions are not
recursive.

A PTA is \emph{(calling) context-insensitive} when it is
interprocedural, but does not distinguish between calls to a function at
different call-sites. For example, a context-insensitive unification-based PTA
would not be able to tell apart \code{str4} and \code{r} passed to \code{print}
in $P_1$, as illustrated in \refsubfig{fig:ov_p1}{fig:p1_steens}. 
A context-insensitive unification-based analysis is obtained by extending
$\ruleSet{Steens}$ with rules for interprocedural assignments.

A \emph{(calling) context-sensitive} PTA provides $\ptsTo$ facts relative to the
requested calling context. In unification-based analyses, this is usually
achieved by calculating a separate $\ptsToInM{F}$ relation for each function $F$ in
the analyzed program. \Dsa is an example of such an analysis \cite{dsa}.
Although not formally specified in~\cite{dsa}, it is defined by
adding rules to $\ruleSet{Steens}$, ${\ruleSet{DSA} = \ruleSet{L} \cup
  \ruleSet{BU} \cup \ruleSet{TD}}$, where ${\ruleSet{L} = \ruleSet{Steens} \cup
  \ruleSet{Formals}}$.

\paragraph{Formal arguments}
To perform a local analysis of a function $F$, \Dsa calculates
$\ptsToIn{F}$ based on instructions in~$F$, including function
calls. These instructions may access memory derived from \emph{formal
  arguments}. Thus, it is necessary to introduce additional abstract
objects for them. We refer to this kind of abstract objects as
\emph{formals}, and provide them for each defined function.  Every
formal argument of a function $i: \code{fun fn(f): r}$, $\code{f}_k$,
has six associated formals: $V_{i, k}^{\fldA}, V_{i, k}^{\fldB}, V_{i, k}^{\fldA
  \fldA}, V_{i, k}^{\fldA \fldB}, V_{i, k}^{\fldB \fldA}, V_{i, k}^{\fldB
  \fldB}$, corresponding to 
abstract objects for fields $\fldA$ and $\fldB$, and abstract objects
reachable by dereferencing each of these two
fields. Fig.~\ref{fig:rules_formals} shows inference rules
$\ruleSet{Formals}$ that specify how these abstract objects may point
to each other. The
rules model precisely only two levels of indirection. Any memory
object obtained by a further dereference is mapped to the same
second-level formal, adding a cycle. In practice, precision of
analysis can be improved by computing the necessary levels of
indirection (e.g.,~\cite{sui-summaries,dsa,sea-dsa}).

\begin{figure}
\begin{mdframed}
\begin{mathpar}
  \squeezeRule  
  \mkRule[Formals]{i: \fnDecl{fn} \\ \mkSpace 0 \leq k < |\code{\cvf}| }
                  {\rfk \ptsToIn{fn} V_{i, k}^{\fldA}
                   \\ \mkSpace V_{i, k}^{\fldA} \ptsToIn{fn} V_{i, k}^{\fldA \fldA}
                   \\ \mkSpace V_{i,k}^{\fldB} \ptsToIn{fn} V_{i,k}^{\fldB \fldA}
                   \\\\ V_{i,k}^{\fldA \fldA} \ptsToIn{fn} V_{i,k}^{\fldA \fldA}
                   \\ \mkSpace V_{i,k}^{\fldB \fldA} \ptsToIn{fn} V_{i,k}^{\fldB \fldA}
                   \\\\ V_{i,k}^{\fldA \fldB} \ptsToIn{fn} V_{i,k}^{\fldA \fldB}
                   \\ \mkSpace V_{i,k}^{\fldB \fldB} \ptsToIn{fn} V_{i,k}^{\fldB \fldB}} 
\end{mathpar}
\end{mdframed}
\caption{Inference rules for formal arguments: $\ruleSet{Formals}$. \label{fig:rules_formals}}
\end{figure}                 

\paragraph{Oversharing}
While the local analysis $\ruleSet{L}$ only uses abstract objects from 
the analyzed function (i.e., coming from allocation sites in that function or
its formals), the rules $\ruleSet{BU} \cup \ruleSet{TD}$, shown in
Fig.~\ref{fig:rules_cs}, propagate $\ptsTo$ facts across functions. They use a
helper function $\mkFn{Resolve}$ to map between caller and callee abstract
objects. For any pair of functions $F_1$ and $F_2$, we refer to
the abstract objects defined by $F_2$ and present in $\ptsToInM{F_1}$ as
\emph{foreign}. A foreign object is \emph{overshared} in $F_1$ if it is
inaccessible by $F_1$, but needlessly appears in the analysis results of $F_1$.

\Dsa, as presented in \cite{dsa}, executes three phases of the analysis for a
function $F$ as follows: (a) \textsc{Local} phase for $F$; (b)
\textsc{Bottom-Up} for each callee of $F$; and (c) \textsc{Top-Down} for each
caller of $F$. This is equivalent to applying the $\ruleSet{L}$ and
$\ruleSet{BU}$ rules until saturation in a reverse-topological call-graph order,
followed by $\ruleSet{TD}$ in a topological order until saturation. The rules
can be soundly applied in this sequence and no new $\ptsToInM{F}$ facts can be
derived by running any of the phases again. The original \Dsa implementation
performs foreign object propagation during both \textsc{Bottom-Up} and
\textsc{Top-Down}: \textsc{Bottom-Up} copies foreign abstract objects
\emph{accessible} from formal arguments and returned values from a callee to its
callers, while \textsc{Top-Down} copies \emph{all} abstract objects
\emph{accessible} (directly or transitively) from function parameters (actual
arguments) in a caller to its callees, even if they are unused. We notice that
the copying of foreign objects in \textsc{Top-Down}, required to maintain
(\textbf{I2}) from Sec.~\ref{sec:overview}, is a major source of oversharing in
\Dsa. This form of oversharing led to a workaround in~\cite{dsa} that improves
performance at expense of precision by treating all global variables (major
source of foreign objects) context-insensitively.

Our first contribution is to show that such an oversharing of foreign abstract
objects is unnecessary. All abstract objects of a function are known after
\textsc{Local} and \textsc{Bottom-Up} phases:
\vspace{-0.15in}
\begin{theorem}
\label{thm:bu_abstract}
  $( \ruleSet{DSA}\; \vdash_{P}\; x \ptsToInM{F} H ) \implies
  \exists y \cdot ( \ruleSet{L} \cup \ruleSet{BU} \; \vdash_{P}\; y \ptsToInM{F} H )
$, where $x$ and $y$ are registers or abstract objects.
\end{theorem}
\vspace{-0.05in} Theorem~\ref{thm:bu_abstract} states that no new foreign
objects are ever introduced by $\ruleSet{TD}$. The derivable $\ptsToInM{F}$
facts are always over abstract objects \emph{resolved} from callee's abstract
object to caller's abstract objects. The proof of Theorem~\ref{thm:bu_abstract}
follows from the fact that $\ruleSet{L}$ models the operational semantics of
our language, and that our formulation of interprocedural rules explicitly uses
the callee-caller resolution of abstract objects\footnote{Proofs of all of the
theorems are available in the extended version of the paper \cite{sea-thesis}.}.

The simplicity of Theorem~\ref{thm:bu_abstract} is solely due to our
new formulation of \Dsa. Prior works (\cite{dsa,otherdsa}) miss this,
now obvious fact. With our formulation, it is clear that the role of
\textsc{Top-Down} is to use $\ptsToInM{F}$ at a call-site and use it
to instantiate a fully-general summary for a callee by introducing
necessary $\ptsToInM{F}$ between function arguments and formals. If a
client of a PTA requires to know not only $\ptsToInM{F}$ but also all
the mapping from formals to allocation sites each formal may originate
from, it is possible to maintain such information separately, without
introducing oversharing during \textsc{Top-Down}. Our evaluation
(Sec.~\ref{sec:evaluation}) demonstrates that this improves 
performance and precision.

\paragraph{Partial Flow-sensitivity}
Our second contribution is to identify additional opportunities to
reduce oversharing by increasing the precision of the analysis at
interprocedural assignments -- call- and return-sites. Overall
precision of a PTA can be improved by making the \textsc{Local} phase more
precise, or by not propagating the local imprecision
interprocedurally. In \Dsa, a function with an
instruction that operates on two abstract objects can cause these
abstract objects to be grouped in any subsequent function, provided
enough interprocedural assignments. The source of the problem is that
\Dsa preserves any local grouping of abstract objects by maintaining
(\textbf{I2}) and (\textbf{I3}) from Sec.~\ref{sec:overview}. Due to
the $\ruleSet{U}$ rules, such confusion can reduce the
precision of the whole PTA.  For example, once $o_1$ and $o_2$ are
grouped together in $\code{getStr}$ from
\refsubfig{fig:ov_p1}{fig:p1_listing}, in \Dsa the grouping is
propagated bottom-up to $\code{foo}$ and $\code{bar}$.

Flow-sensitivity is a simple way to increase precision of a PTA at a cost of
performance. A flow-sensitive analysis computes a relation $\ptsToInAtM{F}{i}$
not only at the function level~($F$), but also relative to a particular
instruction~($i$) within $F$. To improve precision for interprocedural
assignments, we need to know where each function parameter points to at a
particular call- or return-site. For example, in $P_1$ from
\refsubfig{fig:ov_p1}{fig:p1_listing}, ${\code{str1} \ptsToInAt{getStr}{11}
  o_1}$ at the return statement. We call this refinement \emph{partial
  flow-sensitivity}. We present a set of rules, $\ruleSet{PFS}$ in
Fig.~\ref{fig:rules_fs}, that combine together with $\ruleSet{DSA}$ to define an
analysis called $\ruleSet{PFS-DSA}$. Note that $\ruleSet{PFS}$ \emph{replaces}
the corresponding two rules from $\ruleSet{DSA}$. We assume that 
$\ptsToInAtM{F}{i}$ is externally defined and is a (sound) subset of
$\ptsToInM{F}$. \textsc{Bottom-Up-1} rule of $\ruleSet{PFS}$
propagates $\ptsToInAtM{\code{callee}}{j}$ (points-to information at the
return-site) into $\ptsToIn{caller}$, by \emph{resolving} abstract objects
across these two functions; formals from $\code{callee}$ get matched with
abstract objects passed into it at the call-site, while allocation sites from
$\callee$ are resolved to themselves. Similarly, $\textsc{Top-Down-1}$ 
resolves abstract objects reachable from parameters at a call-site into
\mbox{appropriate formals for the $\code{callee}$.}

Partial flow-sensitivity is much cheaper
than a (full) flow-sensitivity, as we do not even need to maintain a
separate flow-sensitive $\ptsToInM{F}$ at call and return sites. This
is because it is often enough to perform a very cheap local reasoning to
determine that given a local fact $p \ptsToInM{F} o$, $p \not
\ptsToInAtM{F}{i} o$. For instance, $\code{str1} \not
\ptsToInAt{getStr}{11} o_2$ because the variable name $\code{str1}$ is
used explicitly at the return-site, and the variable $\code{str1}$ is
never reassigned, it must only point to $o_1$ at line~$11$.

The only difference between $\ruleSet{PDF-DSA}$ and $\ruleSet{DSA}$ is the use of
the $\ptsToInAtM{F}{i}$ relation instead of $\ptsToInM{F}$ in \textsc{Bottom-Up}
and \textsc{Top-Down} rules, where $\ptsToInAtM{F}{i}$ is a subset of
$\ptsToInM{F}$. Assuming $\ptsToInAtM{F}{i}$ is sound at a call-site
(return-site), every $\ptsToInM{F}$ fact is correctly propagated by the
interprocedural assignment rules.

\begin{figure*}
\begin{mdframed}
\begin{mathpar}
  \mkRule[Bottom-Up-1]{i: \code{\cvy = callee(\cvx)} \\ \mkSpace \fnOf{i} = \caller
                     \\ \mkSpace j: \code{return \cvz}
                     \\\\ \fnOf{j} = \callee \\ \mkSpace \rzk \ptsToIn{callee} H
                     \\ \mkSpace \resolve(i, H, I)} 
                     {\ryk \ptsToIn{caller} I}

\mkRule[Bottom-Up-2]{i: \code{\cvy = callee(\cvx)} \\ \mkSpace \fnOf{i} = \caller
                   \\ \mkSpace \accessible(\callee, J)
                   \\\\ J \ptsToIn{callee} K \\ \mkSpace \resolve(i, J, H)
                   \\ \mkSpace \resolve(i, K, I)} 
                    {H \ptsToIn{caller} I}

\mkRule[Top-Down-1]{i: \code{\cvy = callee(\cvx)} \\ \mkSpace \fnOf{i} = \caller
                  \\ \mkSpace \rxk \ptsToIn{caller} H
                  \\\\ j: \fnDecl{callee} \\ \mkSpace \resolve(i, I, H)} 
                   {\rfk \ptsToIn{callee} I}

  \mkRule[Top-Down-2]{i: \code{\cvy = callee(\cvx)} \\ \mkSpace \fnOf{i} = \caller
                    \\ \mkSpace H \ptsToIn{caller} I
                    \\\\ \resolve(i, J, H) \\ \mkSpace \isFormal{J}
                    \\ \mkSpace \resolve(i, K, I) \\ \mkSpace \isFormal{K}} 
                   {J \ptsToIn{callee} K}

\end{mathpar}
\end{mdframed}
\caption{Inference rules for Context-Sensitivity: $\ruleSet{BU}$ and $\ruleSet{TD}$. \label{fig:rules_cs}}
\end{figure*}

\begin{figure}
\begin{mdframed}
 \begin{mathpar}
  \mkRule[Bottom-Up-1]{i: \code{\cvy = callee(\cvx)} \\ \mkSpace \fnOf{i} = \caller
                        \\ \mkSpace j: \code{return \cvz}
                        \\ \mkSpace \fnOf{j} = \callee \\ \mkSpace \rzk \ptsToInAt{callee}{j} H
                        \\ \mkSpace \resolve(i, H, I)} 
                        {\ryk \ptsToIn{caller} I}
  
  \mkRule[Top-Down-1]{i: \code{\cvy = callee(\cvx)} \\ \mkSpace \fnOf{i} = \caller
                       \\\\ \rxk \ptsToInAt{caller}{i} H
                       \\ \mkSpace j: \fnDecl{callee} \\ \mkSpace \resolve(i, I, H)} 
                      {\rfk \ptsToIn{callee} I}

\end{mathpar}
\end{mdframed}
\caption{Inference rules for Partial Flow-Sensitivity: $\ruleSet{PFS}$. \label{fig:rules_fs}}
\end{figure}

 \section{Be Aware of Your Type}
\label{sec:types}

\newcommand{\sqleq}{\sqsubseteq}

The \emph{effective type rules} of the C11 standard~\cite[Sec.~6.5]{ISO:2011:IIIb} say
that memory is dynamically strongly typed: roughly, a memory read
($\code{load}$) of an object $\mathit{co}$ is valid only when the last write
($\code{store}$) to $\mathit{co}$ was of a compatible type. Thus,
pointers of incompatible types do not alias. 
Other languages, including \Cpp and \textsc{Swift}, impose similar rules
typically called \emph{strict aliasing}. Strict aliasing is widely exploited
in all major optimizing compilers. In this paper, we use it to improve precision
\mbox{of the \textsc{Local} phase of \TeaDsa.}

We assume that a \emph{type compatibility}
relation, $\sqleq$, on types, is provided as an \emph{input} to our
analysis. For our simple language, the 
compatibility relation is defined \mbox{as a \emph{partial order} s.t.:}
\begin{align*}
  \forall \tau \in T \cdot \tau &\sqleq \tyChar{} &
  \forall \tau \in PT \cdot \tau &\sqleq \tyChar{*}
\end{align*}
That is, $\tyChar{}$ is compatible with all other types,
$\tyChar{*}$ is compatible with all pointer types, and every type is
compatible with itself, but  $\tyInt{}$ and $\tyFloat{}$ are not
compatible. In our implementation, we use a
more sophisticated type lattice to handle \llvm's
structure types. It is also possible to use the type lattice
of a compiler frontend (e.g.,~\clang's TBAA tags).

Due to the low-level nature of our language, allocations and function
definitions do not specify the types of objects. To allow untyped allocations
and function arguments, we extend our notion of abstract objects to include
object type. For example, an $i: \code{r = alloc()}$ instruction has ${|fld|
\times |T|}$ allocation sites of a form $H_i^T$ -- one for each field of any
possible type. Similarly, each formal function argument has ${|\{\code{a},
\code{b}, \code{aa}, \code{ab}, \code{ba}, \code{bb}\}| \times |T|}$ formals. 
In our implementation, we discover abstract object types on demand. We modify
the basic $\ptsTo$ relation to include the type of the pointed-to abstract
object and disambiguate it from abstract objects of other types. For example, a
fact $\rr \ptsToTy{T} H$ means: the register $\rr$ may point to the abstract
object $H$ of type $T$. A sample points-to graph for a type-aware PTA of a 
program in Fig.~\ref{fig:p2_listing} is shown in Fig.~\ref{fig:p2_teadsa}.

Although the type of each register is known statically, we only require memory
operations (\code{load} and \code{store}) to access objects using compatible
types, while types used in function calls, \code{cast}, and \code{gep}
instructions are ignored. Instead of relying on declared types, we discover them
at memory accesses, as shown in type-awareness rules $\ruleSet{Ty}$ in
Fig.~\ref{fig:rules_types}. We say that a pointer returned by an $\code{alloc}$
may point to \emph{any} abstract object for the field $\fldA$ created at this
allocation site, and express that with a $\ptsToTy{\tyChar{}}$ fact, as
$\tyChar{}$ is compatible with all types. A $\code{load}$ accesses only the
abstract object pointed-to by the pointer operand if they are of a compatible
type. Similarly, the type of the destination register of a $\code{store}$
dictates which abstract object may be written to. For example, consider a simple
class hierarchy $C \sqleq B \sqleq A$, where $A$ is a \emph{superclass} of both
$B$ and $C$, while $B$ is a \emph{superclass} of $C$. In our formalization, a
${\code{load B* } p}$ can access both abstract objects of type $A$ and $B$,
whereas a ${\code{store } v \code{, B* } p}$ writes to abstract objects of type
$B$ and $C$. Such a conservative handling of memory operations, consistent with
the \emph{strict aliasing rules} of C11, guarantees soundness of a PTA extended
with type-awareness rules.

The $\ruleSet{Ty}$ rules \emph{replace} the rules in $\ruleSet{I}$; we omit
the remaining replacement rules that use $\ptsToTy{T}$ instead of $\ptsTo$, as
the modification is straightforward.
Finally, we define $\ruleSet{TeaDsa}$ to be the modified set rules
$\ruleSet{PFS-DSA}$ based on $\ruleSet{Ty}$ and the $\ptsToTy{T}$ relation.
Type-awareness improves both the local and global analysis precision,
and in turn further reduces oversharing:
\vspace{-0.1in}
\begin{theorem}
  ${\ruleSet{PFS-DSA} \; \vdash_P x \not \ptsTo H \implies \ruleSet{TeaDsa} \;
  \vdash_P x \not \ptsToTy{T} H}$
  \label{thm:types_precision}
\end{theorem}
\vspace{-0.2in}
\noindent
Theorem~\ref{thm:types_precision} says that $\ruleSet{TeaDsa}$ is not less precise than
$\ruleSet{PFS-DSA}$, i.e., no points-to relation not present in analysis results
for $\ruleSet{PFS-DSA}$ is present in analysis results for $\ruleSet{TeaDsa}$.
This is because the type-aware rules for $\code{load}$ and $\code{store}$ 
are similar to $\ruleSet{I}$, except that
they prevent loads from deriving facts about stores of incompatible types. With the
most conservative compatibility relation (i.e., all types are
compatible), the $\ruleSet{Ty}$ would derive exactly the same $\ptsTo$ facts as
$\ruleSet{I}$.
 
\begin{figure}
\begin{mdframed}
\begin{mathpar}
  \hspace*{-1.45em}
  \mkRule[Alloc]{i: \code{r = alloc()}}{\rr \ptsToTy{\code{char}} H_i}

  \hspace*{-0.4em}
  \mkRule[Load]{\code{r = load T* p} \\\\ \rp \ptsToTy{\tyU} H
                \\ \mkSpace \tyT \sqleq \tyU
                \\\\ H \ptsToTy{\tyX} I}{\rr \ptsToTy{\tyX} I}
  
  \hspace*{-0.3em}
  \mkRule[Store]{\code{store r, T* p} \\\\ \rp \ptsToTy{\tyU} H
                 \\ \mkSpace \tyU \sqleq \tyT
                 \\\\ \rr \ptsToTy{\tyX} I}{H \ptsToTy{\tyX} I}
\end{mathpar} 
\end{mdframed}
\caption{Type-awareness rules: $\ruleSet{Ty}$. \label{fig:rules_types}}
\end{figure}

 \section{Implementation and Evaluation}
\label{sec:evaluation}

In this section, we describe our implementation of \TeaDsa and compare its
scalability and precision against other state-of-the-art PTAs. To meaningfully
compare precision, we developed a checker for a class of memory safety
violations, and use it to evaluate the PTAs on a set of C and \Cpp programs. Our
implementation, benchmarks, and experiments are available at
\url{https://github.com/seahorn/sea-dsa/releases/tag/tea-dsa-fmcad19}.

\paragraph{Implementation}
We implemented \TeaDsa on top of \SeaDsa\xspace -- a context-, field-, and
array-sensitive \Dsa-style PTA for \llvm~\cite{sea-dsa}. Our implementation
inherits many of the advantages of \SeaDsa, including: an effective
representation of $\ptsTo$ using a union-find data-structure; three analysis
passes (local, bottom-up, top-down); modular analysis of each function; support
for \code{gep} instructions with fixed and symbolic offsets; handling recursion
by losing context sensitivity for strongly connected components in the call
graph; and, on-demand discovery of abstract objects for fields, formals, as well
as their corresponding types. In the evaluation, we devirtualize indirect calls.
For partial flow-sensitivity, we disambiguate pointers that must alias known
allocation sites from other objects in their points-to sets, and do not
propagate stack-allocated abstract objects bottom-up. We use the type
compatibility relation $\sqleq$ based on the type tags in the code such that the
type of each structure is the same as the type of its first (innermost) field.
Two types are compatible if they have \mbox{the same type tag.}

\paragraph{The client}
We chose a problem of statically detecting \emph{field overflow}
bugs. A field-overflow happens when an instruction accesses a
nonexistent field of an object, such that the memory access is outside
of the allocated memory object. For example, consider the field access
in line~19 in Fig.~\ref{fig:p2_listing} -- loading the value of the
field $\code{d}$ is not safe if the pointer $\code{ie}$ is pointing to
an object of an insufficient size, e.g, $o_6$. To determine whether a
field access through a pointer $p$ causes a field overflow, we
identify the set $A$ of all the allocation sites that $p$ might point
to. Then, any allocation site $a \in A$ of an insufficient size might
cause a field overflow. We have implemented such a
field-overflow-checker in \seahorn~\cite{seahorn}.

\paragraph{Evaluation}
We compare \TeaDsa with two state-of-the-art interprocedural PTAs for
\llvm: \SVF~\cite{svf} and \SeaDsa~\cite{sea-dsa}. \SVF~\cite{svf} is
a flow-sensitive, context-insensitive, inclusion-based PTA. We compare
against two variants of \SVF: the most precise Sparse Flow-sensitive
analysis (\textit{SVF Sparse}), and the same analysis with the
\textit{Wave Diff} pre-analysis. As for \Dsa-style analyses, we use
\textsc{SeaDsa}, \textsc{PFS-SeaDsa}, and \textsc{TeaDsa} to denote
\SeaDsa, our implementation of $\ruleSet{PFS-DSA}$, and our
implementation of $\ruleSet{TeaDsa}$, respectively. Note that we do
not use the \Dsa implementation from \llvm's \textsc{Pool-Alloc}, as
it is not maintained and crashes on \mbox{many of our examples}.

We perform the evaluation on a set of C and \Cpp programs. 
The programs vary in size, ranging from
140kB to 158MB of \llvm bitcode. 
All experiments are done on a Linux machine with two Intel Xeon E5-2690v2
10-core processors and 128GB of memory. We present performance results of
running PTAs in Table~\ref{table:perf} and precision on the field-overflow
detection in Table~\ref{table:prec}. In the tables, \textsc{--} denotes that an
experiment did not finish within 3 hours or exceeded the 80GB memory limit and
was terminated.
To ensure that all PTAs are working in a consistent environment, we
modified \SVF to use the same notion of allocation sites that is used
by \TeaDsa. We asses the precision of the PTAs using our
field-overflow checker. In Table~\ref{table:prec}, we use
\emph{Aliases} to denote the number of reported ${\langle
  \text{allocation site}, \text{accessed pointer} \rangle}$ pairs, and
\emph{Checks} as the number of assertions necessary to show that the
analyzed program is free of field overflow bugs. The lower the
numbers, the more precise a PTA is.

In our experiments, \TeaDsa is almost always the most scalable PTA,
both in terms of runtime and memory use, closely followed by
\textsc{PFS-SeaDsa}. These two analyses scaled an order of magnitude
better than the plain version of \SeaDsa. \TeaDsa
was faster than \SVF, especially on large programs like \llvm tools
(prefix \textsf{llvm-}), where it finished in seconds instead of
hours. 
As for precision, \TeaDsa and \SVF achieved similar results on
most of the smaller programs. \TeaDsa is strictly more precise than
\SeaDsa, and, surprisingly, more precise than \SVF on \Cpp programs
such as \textsf{cass}, \mbox{\textsc{Webassembly}} tools (prefix
\textsf{wasm-}), \llvm tools, and on the C program \textsf{htop} that
uses a \Cpp-like coding style.
When performing a closer comparison of \textsc{PFS-SeaDsa} vs
\textsc{SeaDsa}, we noticed that the performance improvement  can be attributed
to not copying foreign objects during \textsc{Top-Down} (up to $96\%$ shorter
running time on \textsf{wasm-opt}), while partial flow-sensitivity
explains most of the increase in precision (up to  $25\%$ fewer aliases on
\textsf{h264ref}).

\newcolumntype{L}[1]{>{\raggedright\let\newline\\\arraybackslash\hspace{0pt}}m{#1}}
\newcolumntype{C}[1]{>{\centering\let\newline\\\arraybackslash\hspace{0pt}}m{#1}}
\newcolumntype{R}[1]{>{\raggedleft\let\newline\\\arraybackslash\hspace{0pt}}m{#1}}

\newcolumntype{M}{>{\raggedleft\let\newline\\\arraybackslash\hspace{0pt}}m{3.6em}}

\begin{table*}[t]
\footnotesize
\renewcommand{\arraystretch}{1.2}
\centering

\begin{tabular}{|L{4.5em}|R{3.3em}|M|M||M|M||M|M||M|M||M|M|}
\hline
\multirow{3}{*}{\textbf{Program}} & \multirow{3}{3.3em}{\textbf{Bitcode Size [kB]}} & \multicolumn{10}{c|}{\textbf{Results}} \\
  \cline{3-12}
  & & \multicolumn{2}{c||}{Wave Diff} & \multicolumn{2}{c||}{SVF Sparse} & \multicolumn{2}{c||}{\textsc{SeaDsa}} &
  \multicolumn{2}{c||}{\textsc{PFS-SeaDsa}} & \multicolumn{2}{c|}{\TeaDsa} \\
  \cline{3-12}
    & & Runtime [s] & Memory [MB] & Runtime [s] & Memory [MB] & Runtime [s] & Memory [MB] & Runtime [s] & Memory [MB] & Runtime [s] & Memory [MB]\\
\hline
\textsf{sqlite} & 140 & \textbf{$<$1} & 106 & \textbf{$<$1} & 135 & \textbf{$<$1} & 57 & \textbf{$<$1} & \textbf{19} & \textbf{$<$1} & \textbf{19} \\
\hline
\textsf{bftpd} & 268 & \textbf{$<$1} & 114 & \textbf{$<$1} & 133 & \textbf{$<$1} & 50 & \textbf{$<$1} & \textbf{24} & \textbf{$<$1} & \textbf{24} \\
\hline
\textsf{htop} & 320 & \textbf{$<$1} & 216 & 7 & 483 & \textbf{$<$1} & 242 & \textbf{$<$1} & 49 & \textbf{$<$1} & \textbf{37} \\
\hline
\textsf{cass} & 1,384 & \textbf{$<$1} & 399 & 1 & 453 & 1 & 375 & \textbf{$<$1} & \textbf{45} & \textbf{$<$1} & 46 \\
\hline
\textsf{wasm-dis} & 1,420 & 6 & 1,041 & 122 & 4,594 & 1 & 618 & \textbf{$<$1} & 169 & \textbf{$<$1} & \textbf{98} \\
\hline
\textsf{openssl} & 1,504 & 1 & 706 & 2 & 792 & 1 & 683 & \textbf{$<$1} & 59 & \textbf{$<$1} & \textbf{58} \\
\hline
\textsf{wasm-as} & 1,824 & 10 & 1,428 & 195 & 8,249 & 2 & 1,162 & \textbf{$<$1} & 248 & \textbf{$<$1} & \textbf{149} \\
\hline
\textsf{h264ref} & 2,468 & 6 & 1,655 & 7 & 1,784 & 5 & 2,323 & \textbf{$<$1} & \textbf{183} & \textbf{$<$1} & 197 \\
\hline 
\textsf{tmux} & 2,996 & 1 & 586 & 3 & 696 & 1 & 649 & \textbf{$<$1} & 144 & \textbf{$<$1} & \textbf{125} \\
\hline
\textsf{wasm-opt} & 3,520 & 36 & 2,784 & 960 & 33,138 & 51 & 23,339 & \textbf{1} & 1,507 & \textbf{1} & \textbf{308} \\
\hline
\textsf{llvm-dis} & 11,232 & 1,640 & 9,964 & -- & -- & -- & -- & 18 & 4,587 & \textbf{16} & \textbf{3,254} \\
\hline
\textsf{llvm-as} & 14,012 & 4,892 & 15,377 & -- & -- & -- & -- & 24 & 7,130 & \textbf{19} & \textbf{4,100} \\
\hline
\textsf{llvm-opt} & 16,012 & 9,104 & 19,633 & -- & -- & -- & -- & 55 & 20,555 & \textbf{27} & \textbf{8,319} \\
\hline
\textsf{rippled} & 157,804 & -- & -- & -- & -- & -- & -- & 379 & 55,691 & \textbf{308} & \textbf{25,626} \\
\hline
\end{tabular}
\caption{Performance of different PTAs. \label{table:perf}}
\end{table*}

\newcolumntype{N}{>{\raggedleft\let\newline\\\arraybackslash\hspace{0pt}}m{3.8em}}

\begin{table*}[t]
\footnotesize
\renewcommand{\arraystretch}{1.2}
\centering

\begin{tabular}{|L{4.5em}|R{3.3em}|M|M||M|M||M|M||N|N||M|M|}
\hline
\multirow{3}{*}{\textbf{Program}} & \multirow{3}{3.3em}{\textbf{Bitcode Size [kB]}} & \multicolumn{10}{c|}{\textbf{Results}} \\
  \cline{3-12}
  & & \multicolumn{2}{c||}{Wave Diff} & \multicolumn{2}{c||}{SVF Sparse} & \multicolumn{2}{c||}{\textsc{SeaDsa}} &
  \multicolumn{2}{c||}{\textsc{PFS-SeaDsa}} & \multicolumn{2}{c|}{\TeaDsa} \\
  \cline{3-12}
    & & Checks & Aliases & Checks & Aliases & Checks & Aliases & Checks & Aliases & Checks & Aliases\\
\hline
\textsf{sqlite} & 140 & \textbf{$<$1k} & \textbf{$<$1k} & \textbf{$<$1k} & \textbf{$<$1k} & 1k & 3k & 1k & 3k & 1k & 3k \\
\hline
\textsf{bftpd} & 268 & $<$1k & $<$1k & \textbf{$<$1k} & \textbf{$<$1k} & $<$1k & 1k & $<$1k & 1k & $<$1k & $<$1k \\
\hline
\textsf{htop} & 320 & 24k & 26k & 24k & 26k & 110k & 110k & 109k & 109k & \textbf{9k} & \textbf{11k} \\
\hline
\textsf{cass} & 1,384 & 1k & 7k & 1k & 7k & 12k & 14k & 3k & 12k & \textbf{$<$1k} & \textbf{3k} \\
\hline
\textsf{wasm-dis} & 1,420 & 136k & 253k & 132k & 241k & 616k & 634k & 539k & 558k & \textbf{119k} & \textbf{132k} \\
\hline
\textsf{openssl} & 1,504 & \textbf{$<$1k} & 2k & \textbf{$<$1k} & \textbf{2k} & $<$1k & 4k & $<$1k & 4k & $<$1k & 4k \\
\hline
\textsf{wasm-as} & 1,824 & 248k & 424k & \textbf{243k} & 412k & 933k & 957k & 823k & 849k & 293k & \textbf{317k} \\
\hline
\textsf{h264ref} & 2,468 & \textbf{$<$1k} & 38k & \textbf{$<$1k} & 37k & 21k & 174k & 15k & 148k & 3k & \textbf{34k} \\
\hline
\textsf{tmux} & 2,996 & 8k & 17k & \textbf{8k} & \textbf{17k} & 403k & 422k & 391k & 410k & 333k & 350k \\
\hline
\textsf{wasm-opt} & 3,520 & 724k & 1,196k & 718k & 1,174k & 8,851k & 8,637k & 7,632k & 7,466k & \textbf{603k} & \textbf{645k} \\
\hline
\textsf{llvm-dis} & 11,232 & 6,107k & 6,842k & -- & -- & -- & -- & 4,358k & 4,391k & \textbf{1,097k} & \textbf{1,404k} \\
\hline
\textsf{llvm-as} & 14,012 & 12,198k & 13,866k & -- & -- & -- & -- & 8,992k & 9,017k & \textbf{2,138k} & \textbf{2,470k} \\
\hline
\textsf{llvm-opt} & 16,012 & 16,346k & 17,140k & -- & -- & -- & -- & 47,174k & 47,421k & \textbf{9,551k} & \textbf{13,878k} \\
\hline
\textsf{rippled} & 157,804 & -- & -- & -- & -- & -- & -- & 130,957k & 129,910k & \textbf{47,415k} & \textbf{47,848k} \\
\hline
\end{tabular}
\caption{Precision of different PTAs. \label{table:prec}}
\end{table*}

 \section{Related Work}
\label{sec:related}

There is a large body of work on points-to analysis, both for low-level
languages and for higher-level languages like Java. Throughout the paper, we
compare with the closest related work: \Dsa~\cite{dsa} and
\SeaDsa~\cite{sea-dsa}. In Sec.~\ref{sec:evaluation}, we compared empirically
with two context-insensitive, inclusion-based implementations of  
\SVF\cite{svf} -- a state-of-the-art PTA framework for LLVM. In the rest of this
section, we compare with other related
works. 

Sui et al.~\cite{sui-summaries} present a
context-sensitive, inclusion-based pointer analysis, called
\icon. 
The fact that
\icon is an inclusion-based PTA and \SeaDsa is unification-based makes
it hard to compare them without an experimental evaluation.
Unfortunately, \icon is not part of the \SVF framework and its
implementation is not publicly available. Therefore, comparing
experimentally is not possible.

The precision of inclusion-based pointer analyses can be improved by
flow-sensitivity
(e.g.~\cite{HardekopfL11,SuiDX16}). However, unification-based PTA are
always flow-insensitive to retain their efficiency. In
our work, we improve a context-sensitive, unification-based PTA by
making it flow-sensitive only at call and return statements. This
allows us to improve the precision of the analysis without
jeopardizing its efficiency.

Using types to improve precision of a PTA is not new.
Structure-sensitive PTA~\cite{BalatsourasS16} extends a whole-program,
inclusion-based PTA with types. The analysis is object and
type-sensitive (\cite{pick-your-contexts-well}).
This work is orthogonal to ours. The main purpose of type sensitivity
is to distinguish multiple abstract memory objects from a given
(untyped) heap allocation (e.g., \texttt{malloc}) based on their
uses. This avoids aliasing among objects that are originated from the
same allocation wrapper or a factory method. We do not tackle this
problem. Instead, we use types to avoid unrealized aliasing under the
strict aliasing rules. We mitigate the problem of using allocation
wrappers by inlining memory allocating functions.

Rakamaric and Hu~\cite{RakamaricH09} use DSA ability to track types for
an efficient encoding of verification conditions (VC) for program analysis. 
Their approach differs significantly from ours. They do not tackle the
problem of improving the precision of a pointer analysis using
types. Instead, they extract useful type information from a PTA to
produce more efficient VCs.

 \section{Conclusion}
\label{sec:concl}

We identify a major deficiency of context-sensitive unification-based
PTA's, called \emph{oversharing}, that affects both scalability and precision.
We present \TeaDsa -- a \Dsa-style PTA that eliminates a class of oversharing
during the \textsc{Top-Down} analysis phase and further reduces it using
flow-sensitivity at call- and return-sites, and typing information. 
Our evaluation shows that avoiding such an oversharing
makes the analysis much faster than \Dsa, as well as more precise than
\Dsa on our program verification problem. The results are very
promising -- \TeaDsa compares favorably against \SVF in
scalability in the presented benchmarks, and sometimes shows even
better precision results.

 \vspace{2em}
\paragraph{Acknowledgments} This material is based upon work supported by US
NSF grants 1528153 and 1817204 and the Office of Naval Research under contract
no. N68335-17-C-0558 and by an Individual Discovery Grant from the Natural
Sciences and Engineering Research Council of Canada. Any opinions, findings and
conclusions or recommendations expressed in this material are those of the
authors and do not necessarily reflect the views of the Office of Naval
Research.

\bibliographystyle{IEEEtran}
\bibliography{biblio}

\begin{thebibliography}{10}
\providecommand{\url}[1]{#1}
\csname url@samestyle\endcsname
\providecommand{\newblock}{\relax}
\providecommand{\bibinfo}[2]{#2}
\providecommand{\BIBentrySTDinterwordspacing}{\spaceskip=0pt\relax}
\providecommand{\BIBentryALTinterwordstretchfactor}{4}
\providecommand{\BIBentryALTinterwordspacing}{\spaceskip=\fontdimen2\font plus
\BIBentryALTinterwordstretchfactor\fontdimen3\font minus
  \fontdimen4\font\relax}
\providecommand{\BIBforeignlanguage}[2]{{%
\expandafter\ifx\csname l@#1\endcsname\relax
\typeout{** WARNING: IEEEtran.bst: No hyphenation pattern has been}%
\typeout{** loaded for the language `#1'. Using the pattern for}%
\typeout{** the default language instead.}%
\else
\language=\csname l@#1\endcsname
\fi
#2}}
\providecommand{\BIBdecl}{\relax}
\BIBdecl

\bibitem{DBLP:conf/icse/YanSCX18}
\BIBentryALTinterwordspacing
H.~Yan, Y.~Sui, S.~Chen, and J.~Xue, ``Spatio-temporal context reduction: a
  pointer-analysis-based static approach for detecting use-after-free
  vulnerabilities,'' in \emph{Proceedings of the 40th International Conference
  on Software Engineering, {ICSE} 2018, Gothenburg, Sweden, May 27 - June 03,
  2018}, M.~Chaudron, I.~Crnkovic, M.~Chechik, and M.~Harman, Eds.\hskip 1em
  plus 0.5em minus 0.4em\relax {ACM}, 2018, pp. 327--337. [Online]. Available:
  \url{http://doi.acm.org/10.1145/3180155.3180178}
\BIBentrySTDinterwordspacing

\bibitem{pinpoint}
Q.~Shi, X.~Xiao, R.~Wu, J.~Zhou, G.~Fan, and C.~Zhang, ``Pinpoint: fast and
  precise sparse value flow analysis for million lines of code,'' in
  \emph{Proceedings of the 39th {ACM} {SIGPLAN} Conference on Programming
  Language Design and Implementation, {PLDI} 2018, Philadelphia, PA, USA, June
  18-22, 2018}, 2018, pp. 693--706.

\bibitem{seahorn}
\BIBentryALTinterwordspacing
A.~Gurfinkel, T.~Kahsai, A.~Komuravelli, and J.~A. Navas, ``The seahorn
  verification framework,'' in \emph{Computer Aided Verification - 27th
  International Conference, {CAV} 2015, San Francisco, CA, USA, July 18-24,
  2015, Proceedings, Part {I}}, ser. Lecture Notes in Computer Science,
  D.~Kroening and C.~S. Pasareanu, Eds., vol. 9206.\hskip 1em plus 0.5em minus
  0.4em\relax Springer, 2015, pp. 343--361. [Online]. Available:
  \url{https://doi.org/10.1007/978-3-319-21690-4_20}
\BIBentrySTDinterwordspacing

\bibitem{introduction-hind}
\BIBentryALTinterwordspacing
M.~Hind, ``Pointer analysis: haven't we solved this problem yet?'' in
  \emph{Proceedings of the 2001 {ACM} {SIGPLAN-SIGSOFT} Workshop on Program
  Analysis For Software Tools and Engineering, PASTE'01, Snowbird, Utah, USA,
  June 18-19, 2001}, J.~Field and G.~Snelting, Eds.\hskip 1em plus 0.5em minus
  0.4em\relax {ACM}, 2001, pp. 54--61. [Online]. Available:
  \url{https://doi.org/10.1145/379605.379665}
\BIBentrySTDinterwordspacing

\bibitem{Andersen1994ProgramAA}
L.~O. Andersen, ``{Program Analysis and Specialization for the C Programming
  Language},'' Ph.D. dissertation, DIKU, University of Copenhagen, 1994.

\bibitem{Steensgaard96AA}
\BIBentryALTinterwordspacing
B.~Steensgaard, ``Points-to analysis in almost linear time,'' in
  \emph{Proceedings of the 23rd ACM SIGPLAN-SIGACT Symposium on Principles of
  Programming Languages}, ser. POPL '96.\hskip 1em plus 0.5em minus 0.4em\relax
  New York, NY, USA: ACM, 1996, pp. 32--41. [Online]. Available:
  \url{http://doi.acm.org/10.1145/237721.237727}
\BIBentrySTDinterwordspacing

\bibitem{dsa}
\BIBentryALTinterwordspacing
C.~Lattner and V.~S. Adve, ``Automatic pool allocation: improving performance
  by controlling data structure layout in the heap,'' in \emph{Proceedings of
  the {ACM} {SIGPLAN} 2005 Conference on Programming Language Design and
  Implementation, Chicago, IL, USA, June 12-15, 2005}, V.~Sarkar and M.~W.
  Hall, Eds.\hskip 1em plus 0.5em minus 0.4em\relax {ACM}, 2005, pp. 129--142.
  [Online]. Available: \url{https://doi.org/10.1145/1065010.1065027}
\BIBentrySTDinterwordspacing

\bibitem{sea-dsa}
\BIBentryALTinterwordspacing
A.~Gurfinkel and J.~A. Navas, ``A context-sensitive memory model for
  verification of {C/C++} programs,'' in \emph{Static Analysis - 24th
  International Symposium, {SAS} 2017, New York, NY, USA, August 30 - September
  1, 2017, Proceedings}, ser. Lecture Notes in Computer Science, F.~Ranzato,
  Ed., vol. 10422.\hskip 1em plus 0.5em minus 0.4em\relax Springer, 2017, pp.
  148--168. [Online]. Available:
  \url{https://doi.org/10.1007/978-3-319-66706-5_8}
\BIBentrySTDinterwordspacing

\bibitem{svcomp19}
\BIBentryALTinterwordspacing
D.~Beyer, ``Automatic verification of {C} and java programs: {SV-COMP} 2019,''
  in \emph{Tools and Algorithms for the Construction and Analysis of Systems -
  25 Years of {TACAS:} TOOLympics, Held as Part of {ETAPS} 2019, Prague, Czech
  Republic, April 6-11, 2019, Proceedings, Part {III}}, ser. Lecture Notes in
  Computer Science, D.~Beyer, M.~Huisman, F.~Kordon, and B.~Steffen, Eds., vol.
  11429.\hskip 1em plus 0.5em minus 0.4em\relax Springer, 2019, pp. 133--155.
  [Online]. Available: \url{https://doi.org/10.1007/978-3-030-17502-3_9}
\BIBentrySTDinterwordspacing

\bibitem{ISO:2011:IIIb}
\BIBentryALTinterwordspacing
{ISO}, \emph{{ISO\slash IEC 9899:2011 Information technology --- Programming
  languages --- C}}.\hskip 1em plus 0.5em minus 0.4em\relax Geneva,
  Switzerland: International Organization for Standardization, Dec. 2011.
  [Online]. Available:
  \url{http://www.iso.org/iso/iso_catalogue/catalogue_tc/catalogue_detail.htm?csnumber=57853}
\BIBentrySTDinterwordspacing

\bibitem{svf}
\BIBentryALTinterwordspacing
Y.~Sui and J.~Xue, ``{SVF:} interprocedural static value-flow analysis in
  {LLVM},'' in \emph{Proceedings of the 25th International Conference on
  Compiler Construction, {CC} 2016, Barcelona, Spain, March 12-18, 2016},
  A.~Zaks and M.~V. Hermenegildo, Eds.\hskip 1em plus 0.5em minus 0.4em\relax
  {ACM}, 2016, pp. 265--266. [Online]. Available:
  \url{http://doi.acm.org/10.1145/2892208.2892235}
\BIBentrySTDinterwordspacing

\bibitem{pointer-analysis-tutorial}
Y.~Smaragdakis, G.~Balatsouras \emph{et~al.}, ``Pointer analysis,''
  \emph{Foundations and Trends{\textregistered} in Programming Languages},
  vol.~2, no.~1, pp. 1--69, 2015.

\bibitem{sui-summaries}
\BIBentryALTinterwordspacing
Y.~Sui, S.~Ye, J.~Xue, and J.~Zhang, ``Making context-sensitive inclusion-based
  pointer analysis practical for compilers using parameterised summarisation,''
  \emph{Softw., Pract. Exper.}, vol.~44, no.~12, pp. 1485--1510, 2014.
  [Online]. Available: \url{https://doi.org/10.1002/spe.2214}
\BIBentrySTDinterwordspacing

\bibitem{sea-thesis}
\BIBentryALTinterwordspacing
J.~Kuderski, ``Scalable context-sensitive pointer analysis for {LLVM},''
  Master's thesis, University of Waterloo, 2019. [Online]. Available:
  \url{https://hdl.handle.net/10012/14875}
\BIBentrySTDinterwordspacing

\bibitem{otherdsa}
\BIBentryALTinterwordspacing
R.~Madhavan, G.~Ramalingam, and K.~Vaswani, ``A framework for efficient modular
  heap analysis,'' \emph{Foundations and Trends in Programming Languages},
  vol.~1, no.~4, pp. 269--381, 2015. [Online]. Available:
  \url{https://doi.org/10.1561/2500000020}
\BIBentrySTDinterwordspacing

\bibitem{HardekopfL11}
B.~Hardekopf and C.~Lin, ``Flow-sensitive pointer analysis for millions of
  lines of code,'' in \emph{Proceedings of the {CGO} 2011, The 9th
  International Symposium on Code Generation and Optimization, Chamonix,
  France, April 2-6, 2011}, 2011, pp. 289--298.

\bibitem{SuiDX16}
Y.~Sui, P.~Di, and J.~Xue, ``Sparse flow-sensitive pointer analysis for
  multithreaded programs,'' in \emph{Proceedings of the 2016 International
  Symposium on Code Generation and Optimization, {CGO} 2016, Barcelona, Spain,
  March 12-18, 2016}, 2016, pp. 160--170.

\bibitem{BalatsourasS16}
\BIBentryALTinterwordspacing
G.~Balatsouras and Y.~Smaragdakis, ``Structure-sensitive points-to analysis for
  {C} and {C++},'' in \emph{Static Analysis - 23rd International Symposium,
  {SAS} 2016, Edinburgh, UK, September 8-10, 2016, Proceedings}, ser. Lecture
  Notes in Computer Science, X.~Rival, Ed., vol. 9837.\hskip 1em plus 0.5em
  minus 0.4em\relax Springer, 2016, pp. 84--104. [Online]. Available:
  \url{https://doi.org/10.1007/978-3-662-53413-7_5}
\BIBentrySTDinterwordspacing

\bibitem{pick-your-contexts-well}
\BIBentryALTinterwordspacing
Y.~Smaragdakis, M.~Bravenboer, and O.~Lhot\'{a}k, ``Pick your contexts well:
  Understanding object-sensitivity,'' in \emph{Proceedings of the 38th Annual
  ACM SIGPLAN-SIGACT Symposium on Principles of Programming Languages}, ser.
  POPL '11.\hskip 1em plus 0.5em minus 0.4em\relax New York, NY, USA: ACM,
  2011, pp. 17--30. [Online]. Available:
  \url{http://doi.acm.org/10.1145/1926385.1926390}
\BIBentrySTDinterwordspacing

\bibitem{RakamaricH09}
\BIBentryALTinterwordspacing
Z.~Rakamaric and A.~J. Hu, ``A scalable memory model for low-level code,'' in
  \emph{Verification, Model Checking, and Abstract Interpretation, 10th
  International Conference, {VMCAI} 2009, Savannah, GA, USA, January 18-20,
  2009. Proceedings}, ser. Lecture Notes in Computer Science, N.~D. Jones and
  M.~M{\"{u}}ller{-}Olm, Eds., vol. 5403.\hskip 1em plus 0.5em minus
  0.4em\relax Springer, 2009, pp. 290--304. [Online]. Available:
  \url{https://doi.org/10.1007/978-3-540-93900-9_24}
\BIBentrySTDinterwordspacing

\end{thebibliography}

\end{document}